\documentstyle[aps,prl,preprint]{revtex}
\begin{document}

\title {A Displacive Structural Transformation in the CuO$_{2}$ Planes
of YBa$_2$Cu$_3$O$_x$ at the Underdoped-Overdoped Phase Separation
Line
}

\author{E. Kaldis$^{1}$, J. R\"{o}hler$^{2}$, E. Liarokapis$^{3}$,
N. Poulakis$^{3}$, K. Conder$^{1}$ and P.W. Loeffen$^{4}$
}

\address {$^{1}$Labor f\"{u}r Festk\"{o}rperphysik, ETH-Technopark,
CH-8005 Z\"{u}rich, Switzerland\\
$^{2}$II. Physikalisches Institut, Universit\"{a}t zu K\"{o}ln, D-50937
K\"{o}ln, Germany
\\$^{3}$National Technical University of Athens, Athens 157 80,
Greece
\\$^{4}$European Synchrotron Radiation Facility, F-38043 Grenoble,
France
}

\date{\today}
\maketitle

\begin{abstract}
A structural phase transformation in the CuO$_2$ planes of
YBa$_2$Cu$_3$O$_x$ has been observed at the onset of the overdoped
regime, $x=6.95$. We have measured as a function of {\em x} the dimpling
in the CuO$_2$ planes by EXAFS, and the O2,3 in-phase $(A_{1g})$ mode by
Raman scattering. The data show for $x\geq 6.95$ anomalously large
static displacements of the Cu2 atoms off the O2,3 layer, and a gap in
the distribution of the O2,3 in-phase Raman shifts. We conclude the
structure of overdoped YBa$_2$Cu$_3$O$_x$ to be a martensitic form of
the structure most favourable for the superconductivity, and possibly at
the origin of the split superconducting transition in the overdoped
regime.
\end{abstract}



The so-called 90 K plateau of YBa$_2$Cu$_3$O$_x$ is well established
\cite{RusTen,ClaLoh,ConJil} to exhibit a broad but clear maximum of
$T_c=92.5$ K at optimum doping around $x_{opt}=6.92$, see Fig.\
\ref{Fig1}(a). In the {\em T-x} phase diagram of the cuprate
superconductors $x_{opt}$ indicates the phase separation line between
the under- and overdoped regimes. While optimum doped
YBa$_2$Cu$_3$O$_{6.92}$ exhibits a single superconducting transition
various laboratories have reported from measurements of the
magnetization \cite{ConJil,ZecMue,Kal97,JanMul},
resistivity\cite{LorMir} and specific heat \cite{JanMul,LorMir} of
overdoped YBa$_2$Cu$_3$O$_x$, $x\geq 6.95$, two superconducting phases
with critical temperatures split by about 2.5 K. Optimum doped
YBa$_2$Cu$_3$O$_{6.92}$ has an intermediate oxygen concentration with
respect to its insulating (antiferromagnetic) parent phase and the
overdoped metallic phase. It is an important indication that only a
very narrow compositional range is most favourable for the
superconductivity. The physical reason for the narrow phase separation
line between the under- and overdoped regimes to occur just at an
intermediate oxygen concentration of $x=6.92$ is a matter of intense
and controversial debates \cite{PS}.

In this Letter we report evidence for a structural phase
transformation in the superconducting CuO$_2$ planes of
YBa$_2$Cu$_3$O$_x$ occuring close to optimum doping at the onset of
the overdoped regime, $x=6.95$. We show a martensitic form of the
crystal to develop across the underdoped--overdoped phase separation
line also in ``classically'' prepared YBa$_2$Cu$_3$O$_x$ exhibiting no
anomalies in the {\em x}\/-dependency of the lattice parameters. A
lattice instability around optimum doping has been earlier suggested
\cite{RusTen,ConJil} from the observation of a minimum in the {\em
c}\/-axis lattice parameter, see the open symbols in Fig.\
\ref{Fig1}(b). However, the experimental evidence came exclusively
from special samples synthesized by direct oxidation of the metals
(DO) or with Ba obtained from decomposed BaCO$_3$ (BaO)
\cite{RusTen,ConJil}. On the other hand hundreds of samples,
``classically'' synthesized by direct reaction of BaCO$_3$ with the
metal oxides (CAR), all exhibited a linear {\em x}\/-dependency of the
{\em c}\/-axis, extending straight across the boundary between the
under- and overdoped regimes \cite{KruKal}, cf. the closed squares in
Fig.\ \ref{Fig1}(b). Therefore the lattice anomalies found in the DO
and BaO samples could be possibly an artefact arising from their
particular form of carbonate incorporation \cite{MacKal}.

The different conditions of synthesis \cite{ConJil,ZecMue,Kal97}, and
the applied method for high precision measurements of the oxygen
concentrations \cite{ConKal} have been reported elsewhere. We
emphazise that all these methods of synthesis used very slow cooling
rates to obtain homogenous samples under near equilibrium conditions.
Originally the oxygen diffusion in YBa$_2$Cu$_3$O$_x$ was assumed to
freeze at $T<500^{\circ}$C, and thus the samples were quenched by
various laboratories\cite{JanMul,CavRup}. Detailed studies of the
oxygen diffusion process\cite{ConKel,ConKru}, however, have
unambigously shown oxygen to diffuse rapidly even at $T<250^{\circ}$C,
and consequently slow cooling is a necessary prerequisite to approach
the equilibrium state of lattice defects. The CAR samples were found
in all laboratories (also when slowly cooled) to exhibit a linear {\em
x}\/-dependency of the {\em c}\/-axis. It is not known at the present
time why the use of BaCO$_3$ as a precursor masks the {\em c}\/-axis
minimum found in the DO samples. We believe that the chemical history
\cite{ConJil,Kal97,MacKal} is as important as the thermal history for
the synthesis of homogenous samples near the equilibrium.

The {\em c}\/-axis contraction upon oxygen doping is usually
attributed to the electron-hole charge transfer to the CuO$_{2}$
planes shortening predominantly the Cu2-O1 apical bond. We may
therefore expect a minimum of $c(x)$ around $x_{opt}$ to correlate
with a minimum in the {\em x}\/-dependency of the apical bond. But
standard refinements of neutron diffraction patterns from the DO
samples have shown that the shortening of the Cu2-O1 apical bond
extends with doping into the overdoped regime, thereby changing very
slightly its slope \cite{ConJil}. A step-like discontinuity at
$x=6.947$ has been found to reduce the dimpling of the CuO$_2$ planes
by $-0.012$ $\rm\AA$ \cite{ConJil}. However, the negative direction of
this discontinuity is hard to reconcile with the inversion of the {\em
c}\/-axis parameter and the internal bondlengths. Most likely the
structural transformation develops first in small domains of the
crystal which are not accessible to standard diffraction techniques.

We have therefore explored the atomic structure of the CuO$_2$ planes
on a nanoscale by extended $x$\/-ray absorption-fine-structure
spectroscopy (EXAFS), and its dynamics by Raman scattering. We have
examined CAR samples from the same batches, all without anomaly in
$c(x)$, cf. Fig.\ \ref{Fig1}(b), closed squares.

The static displacements of the planar Cu2 and O2,3 atoms upon doping
have been determined by yttrium {\em K}\/-edge EXAFS. The photoexcited
Y atoms are ideal observers of the local CuO$_{2}$ structure. In
particular we have exploited the high sensitivity of the nearly
collinear three-body electron scattering configurations Y-Cu2-Ba and
Y-O2,3-Ba for displacements of the intervening Cu2 and O2,3 atoms,
respectively. Since both multiple scattering paths refer to the same
Ba-layer, and show up extremely well isolated in real space, the
dimpling of the CuO$_2$ planes may be directly read from the
magnitudes of the Fourier transform spectra at 5 $\rm\AA$ and 6.2
$\rm\AA$. The spectra have been recorded in transmission geometry from
finely grained polycrystalline absorbers. Details of the experiments
and the data analysis are given elsewhere \cite{RoeKal}. Fig.\
\ref{Fig1}(c) displays the spacings between the Y layer and the O2,3
and Cu2 layers, respectively, as a function of $x$ ($T=25$ K). Their
differences yield the $x$\/-dependency of the dimpling. On doping from
the underdoped side, $6.8<x\rightarrow x_{opt}$, the Cu2 position is
found to move along $c$ by about 0.025 {\rm\AA} off the O2,3 layer
while the average O2,3 positions remain almost unaffected. Around
$x=6.95$ the dimpling increases by another 0.04 {\rm\AA} to its
maximum value of 0.30 $\rm\AA$, nearly entirely due to displacements
of the Cu2 atoms. At $x=6.984$ both, the O2,3 and the Cu2 layers shift
off the Y layer, thereby reducing the dimpling to 0.28 $\rm\AA$.

The Raman shifts, $\bar{\nu}$, of the O2,3 in-phase mode ($\rm
A_{1g}$) plotted in Fig.\ \ref{Fig2} (bottom) have been recorded at
300 K in the scattering configuration $y(zz)\bar{y}$ from a total
number of 97 microcrystallites ($\approx$15 $\mu$m sized), in the
average about 8 at each measured concentration between x=6.438 and
6.984. The {\em x}\/-dependency of the O2,3 in-phase Raman shift at
300 K has been earlier shown to decrease with increasing oxygen
content, and, most important, to soften strongly around $x_{opt}$
\cite{PouMue,Lia}, cf. the diagonal lines in Fig.\ \ref{Fig2}
(bottom). A similar behaviour has been recently found also at 200 K
and 77 K \cite{Lia}. Fig.\ \ref{Fig2} (top) exhibits the normalized
distribution function, $P(\bar{\nu},x)$, of the observed Raman shifts,
$\bar{\nu}$. It is clearly visible that $P(\bar{\nu},x)$ peaks at 6
characteristic wavenumbers, labelled A-F. Therefore $\bar{\nu}(x)$ is
better described by staircases than by straights. In the underdoped
regime we identify single phase regions (F, E, D) alternating with
two-phased regions (F+E, D+E). The single phase regions can be
correlated with the particular stability of the superstructures:
$2a_0$ around $x_{2a}=6.5$ (F), $3a_0$ around $x_{3a}=6.66$ (E), and
$5a_0$ around $x_{5a}=6.8$ (D). Here $a_0$ denotes the {\em
a}\/-lattice parameter of the fundamental unit cell.

For oxygen concentrations $6.8\le x\le 6.89$ we find a multi phase
region composed of the underdoped phase D, the optimum doped phase C
(subdivided into C$_u$ at the underdoped flank, and C$_o$ at the
overdoped flank), and possibly a weak admixture of the overdoped phase
B.

Around optimum doping, $x_{opt}=6.92$, C$_u$ dominates within the
$-0.03$ wide region below, and C$_o$ in the $+0.03$ wide concentration
region above. While the underdoped phase D vanishes completely some
admixtures of the overdoped phase B start to appear. The latter
contribute strongly at the overdoped flank, $x=6.92$--6.975.

The overdoped regime exhibits the phases A and B. Most important,
while for $x<6.95$ all phases transform continously, a frequency gap
at 434 cm$^{-1}$ (drawn out vertical lines) suggests for $x>6.95$ a
miscibility gap between the overdoped phases A and B.

We relate the anomalous softening of the O2,3 Raman shifts and the
miscibility gap in the overdoped regime to the anomalously large
displacements of the Cu2 atoms off the O2,3 layer observed by EXAFS
around $x=6.95$, cf. Fig.\ \ref{Fig1}(c). It is intuitively plausible
that the increase of the dimpling in the CuO$_2$ planes softens the
Cu2-O2,3 bonds, and thus may decrease the wavenumber of the O2,3
in-phase vibrations. The drop of the Raman shift by $-5$ cm$^{-1}$
within an extremely narrow concentration range of $\Delta x\simeq
0.025$ gives strong evidence for that the deformation of the CuO$_2$
planes is of the displacive type. The gap in the Raman shift indicates
a first order transition. From the relatively few data points in Fig.\
\ref{Fig1}(c) evidencing the static anomaly in the dimpling we may
also infer a first order transition, at least of the moderate type.

From the structural data displayed in Fig.\ \ref{Fig1}(b), open
symbols, and Fig.\ \ref{Fig1}(c) we conclude that the {\em
x}\/-dependency of the $c(x)$\/-axis lattice parameter scales
inversely with the dimpling of the CuO$_2$ planes, correlated with the
electron-hole charge transfer along $c$. In the underdoped phase
mixture the dimpling increases with decreasing {\em c}, and exhibits
its maximum value of 0.30 {\rm\AA} close to the {\em c}\/-axis
minimum. In the overdoped regime, $x\geq6.95$, $c(x)$ increases while
the dimpling decreases, but the possible correlation is weaker than in
the underdoped regime. Our findings are in conflict with the
discontinuity in the dimpling extracted from neutron diffraction
patterns of the DO samples \cite{ConJil}. However, it is well known
that the crystallographic structure of martensitic phases, cf. e.g.
Ref.\ \cite{Kru}, is not reliably accessible from standard refinements
of the diffraction data, and consequently further diffraction work is
invoked.

The important contribution of this work in the field of cuprate
superconductivity is in showing that the phase separation line between
the under- and overdoped regimes is accompanied by a structural phase
transformation deforming the CuO$_2$ planes. Whereas the concentration
region around optimum doping is composed of many coexisting
intermediate phases, the overdoped regime exhibits a miscibility gap
and a martensitic structural transformation. It is a widely known fact
that all superconducting compounds including the old intermetallic
high-$T_c$ materials (e.g. A15's) and ``old oxides'' undergo
martensitic structural transformations \cite{Kru,Mat}. In many A15's
double superconducting transitions have been observed, and by applying
strain the lower transition temperature has been shown to arise from
the deformed martensitic phase \cite{Mat}. A large coupling between
atomic displacements and the shape of the superconducting transition
exists also in the heavy fermion compound UPt$_3$ where annealing
treatment of as grown crystals causes the specific heat anomaly to
evolve from a broad bump into a sharp double peak feature
\cite{MidLoe}.

Our results support the idea that the dimpling of the CuO$_2$ planes
is an intrinsic property of the superconducting cuprates, and that
local coherent distortions of the CuO$_2$ planes and the
superconductivity are coupled. We hope that porposals for the driving
mechanism of the reported martensic phase transformation at the
underdoped-overdoped phase separation line of YBa$_2$Cu$_3$O$_x$ might
be available in the near future.

We thank S. M\"{u}llender, and S. Link for help with the EXAFS
experiments, the European Synchrotron Radiation Facility for beamtime
and the use of the BM29 facilities. Work at the ETH has been supported
by the NFP30 of the Swiss Nat. Fonds.

\begin{figure}
\caption{\label{Fig1}
(a) Superconducting transition temperature, $T_c$, of YBa$_2$Cu$_3$O$_x$
around optimum doping, $x_{opt}$, as determined from magnetization
measurements. Vertical lines indicate the boundaries between the phases
analyzed in Fig. 2. (b) {\em c}\/-axis lattice parameter as a function
of oxygen concentration from powder {\em x}\/-ray diffraction, by
courtesy of Chr. Kr\"{u}ger. CAR samples (filled squares), DO samples
(open squares), BaO samples (open diamonds). The superimposed lines are
guides to the eyes. (c) Spacing between the Y-Cu2 and Y-O2,3 layers,
respectively, as determined from Y-EXAFS. Vertical arrows indicate the
dimpling.} \end{figure}

\begin{figure}
\caption{\label{Fig2}
(bottom)\/ Raman shifts of the O2,3 in-phase mode in YBa$_2$Cu$_3$O$_x$
for oxygen concentrations between $x=6.438$ and 6.984. Dashed horizontal
lines indicate the phase boundaries between coexisting phases, the drawn
out horizontal lines (6.95) the miscibility gap in the overdoped regime.
The thick drawn boxes emphazise the sequence of phases occuring on
doping. (top)\/ Distribution function of the Raman shifts normalized to
a constant number of measurements in evenly spaced doping intervals. The
maxima A-F are attributed to the different phases. Peak C labelling the
the optimum doped phases is subdivided into $\rm C_o$ (at the overdoped
side of x$_{opt}$) and $\rm C_u$ (at the underdoped side of x$_{opt}$).
}
\end{figure}
\end{document}